\documentclass[preprint,aps]{revtex4}

\usepackage{graphicx}
\usepackage{dcolumn}

\begin{document}

\preprint{Lebed-CDW-PRL}

\title{Universal Field-Induced Charge-Density-Wave 
Phase Diagram: Theory versus Experiment}

\author{A.G. Lebed$^*$}

\affiliation{Department of Physics, University of Arizona, 1118 E.
4-th Street, Tucson, AZ 85721, USA}

\begin{abstract}

We suggest a theory of the Field-Induced Charge-Density-Wave 
(FICDW) phases, generated by high magnetic fields in quasi-low-dimensional 
conductors.
We demonstrate that, in layered quasi-one-dimensional conductors, 
the corresponding critical magnetic fields ratios are universal and do 
not depend on any fitting parameter.
In particular, we find that $H_1/H_0 = 0.73, \ H_2/H_0 = 0.59, \
H_3/H_0 = 0.49, \ H_4/H_0 = 0.42$, where 
$H_n$ is a critical field of a phase transition between the FICDW phases 
with numbers 
$n$ and $n+1$.
The suggested theory is in very good qualitative and quantitative agreements 
with the existing experimental data in 
$\alpha$-(ET)$_2$KHg(SCN)$_4$ material.
\end{abstract}

\maketitle

\pagebreak

High magnetic field properties of organic conductors and superconductors 
have been intensively studied [1,2] since a discovery of the so-called 
field-induced spin-density-wave (FISDW) phase 
diagrams [3,4].
Phase transitions from metallic to the FISDW phases were successfully 
explained in terms of the $3D \rightarrow 2D$ dimensional crossovers 
[5-11,1].
In particular, the metal-FISDW phase transition line was calculated in Refs. [5-7], 
whereas, a free energy of the FISDW phases was evaluated for all range of temperatures and magnetic fields in 
Refs. [8,9].
In addition, the so-called three-dimensional quantum Hall effect, experimentally 
observed in the FISDW phases [1-4], was theoretically explained in 
Refs. [10,11].

A relative phenomenon - the so-called field-induced charge-density-wave
(FICDW) phase diagram - was anticipated in Refs. [5,12] and recently 
experimentally discovered in $\alpha$-(ET)$_2$KHg(SNC)$_4$ 
conductor [13-18].
Although originally the FICDW phases were predicted to exist due to
electron-electron interactions [12], later it was shown [19] that they naturally 
appeared in a physical picture, where only electon-phonon interactions were
taken into account.
Note that the phase diagram, suggested in Ref. [12], depends on many parameters
such as details of electron-electron interactions, temperature, and anisotropy 
ratios of a quasi-one-dimensional (Q1D) electron 
spectrum.
In addition, according to Ref. [12], the FICDW phases are always mixed with the 
FISDW ones. 
The above mentioned circumstances make it to be almost impossible to test 
the theory [12] and to compare it with the existing experiments 
[13-18].
In a model [19], based on electron-phonon interactions, there is no the 
FICDW-FISDW mixing effects, but the analysis [19] is oversimplified and, as 
we stress below, is not in a quantitative agreement with the 
experimental data.

The main goal of our Letter is to suggest a universal theory of the FICDW
phase diagram, which does not depend on details of electron-electron and
electron-phonon interactions as well as on temperature and details of a Q1D 
electron spectrum.
In particular, we suggest a model, based on electron-phonon interactions,
for a general form of a layered Q1D spectrum. 
We demonstrate that the critical magnetic fields ratios, $H_1/H_0 = 0.73, \ 
H_2/H_0 = 0.59, \ H_3/H_0 = 0.49, \ H_4/H_0 = 0.42$ (where $H_n$ is a critical 
field of a phase transition between the FICDW phases with numbers 
$n$ and $n+1$) do not depend on any parameter and calculate 
them.
A comparison of the present theory with the experiments [14-18] shows not 
only qualitative but also quantitative agreement.
This justifies a validity of our approach and indicates, in particular, that the
electron-electron interactions and FICDW-FISDW mixing effects [12] are 
not very important.

Let us consider the most general layered Q1D electron spectrum, linearized near
its two Fermi surface (FS) sheets,
\begin{eqnarray}
&&\epsilon^{\pm} ({\bf p}) = \pm v_F (p_x \mp p_F) + t^0_y (p_y a_y) 
+ t^0_z (p_z a_z), 
\nonumber\\ 
&&t^0_y (p_y a_y) = 2t_y  \cos (p_y a_y \pm \alpha) , \ \ 
t^0_z (p_z a_z) = 2t_z \cos (p_z a_z \pm \beta) \ , 
\end{eqnarray}
which obeys the so-called "nesting" condition [1,2],
\begin{equation}
\epsilon ({\bf p} + {\bf Q_0}) + \epsilon ({\bf p}) = 0 , \ \
{\bf Q_0} = [2p_F,  (\pi - 2 \alpha)/a_y ,  (\pi - 2 \beta)/a_z ] . 
\end{equation}
[Here $+(-)$ stands for right (left) sheet of Q1D FS (1); $p_F$ and $v_F$ 
are the Fermi momentum and Fermi velocity, respectively; $t_y$ and $t_z$
are overlapping integrals between electron wave functions;  $p_F v_F \gg t_y
\gg t_z$; $\alpha$ and $\beta$ are some phase shifts; 
$\hbar \equiv 1$.]
It is well known [1,2,5-9,12,19] that the so-called Peierls instability for "nested" 
FS  (1) results in the appearance of a density wave 
ground state.
Below, we consider a CDW ground state in accordance
with the existing experimental data in $\alpha$-(ET)$_2$KHg(SCN)$_4$ 
material [13-18].

If we take into account a small (but finite) non-linearity in a Q1D electron spectrum 
(1) along the conducting chains, then we obtain the following electron spectrum,
\begin{eqnarray}
&&\epsilon^{\pm} ({\bf p}) = \pm v_F (p_x \mp p_F) + t_y (p_y a_y) , 
\nonumber\\ 
&&t_y (p_y a_y) = 2t_y \cos (p_y a_y \pm \alpha) + 
2t^{'}_y \cos (2p_y a_y \pm 2 \alpha),  
\end{eqnarray}
with small "antinesting" term, $2t^{'}_y \cos (2p_y a_y \pm 2 \alpha)$, 
where $t^{'}_y \sim t^2_y / (p_F v_F) \ll t_y$.
[Note that, in Eq.(3), we use a 2D model electron spectrum, since we suggest that 
$t_y  \gg t_z$.
In this case, the CDW and FICDW phases always correspond to an ideal "nesting" vector 
(2) along z-axis since the corresponding "antinesting" term is too small, $t^{'}_z
\sim t^2_z / (p_F v_F) \ll t^{'}_y$.]
The "antinesting" term in Eq.(3) is known to decrease a stability of the 
CDW ground state and, therefore, at high pressures (i.e., large enough values of $t'_y$), 
metallic phase has to be restored [1,2,5-9].

At first, let us discuss the FICDW phases formation, using qualitative arguments.
For this purpose, we consider a Q1D electron spectrum (3) in the presence of an external magnetic field, applied along z-axis, 
\begin{equation}
{\bf H} = (0,0,H), \ \ \ {\bf A} = (0,Hx,0) .
\end{equation}
To obtain electron Hamiltonian in a magnetic field (4) from the spectrum (3) we
use the Peierls substitution method, $p_x \rightarrow - i (d / dx)$, 
$p_y \rightarrow p_y - (e/c) A_y$, and take into account the Pauli
spin splitting effects, 
\begin{equation}
\biggl[ \pm v_F \biggl( -i \frac{d}{dx} \mp p_F \biggl) 
+ t_y \biggl( p_y a_y - \frac{\omega_c}{v_F}x \biggl) - \mu_B \sigma H \biggl]
\Psi^{\pm}_{\epsilon}(x,p_y,\sigma) = 
\delta \epsilon \ \Psi^{\pm}_{\epsilon}(x,p_y,\sigma ) ,
\end{equation}
where $\sigma = +1 (-1)$ for spin up (down), $\omega_c = ev_FH a_y /c$,
$\delta \epsilon = \epsilon - \epsilon_F$.

It is important that Eq.(5) can be solved and the corresponding wave functions 
can be determined analytically,
\begin{equation}
\Psi^{\pm}_{\epsilon}(x,p_y,\sigma) =  \exp(\pm i p_F x) 
\exp \biggl( \pm i \frac{\delta \epsilon}{v_F}x \biggl)
\exp \biggl( \pm i \frac{\mu_B \sigma H}{v_F}x \biggl)   
\exp \biggl[ \mp \frac{i}{v_F} \int^x_0 t_y \biggl(p_y a_y - 
\frac{\omega_c}{v_F} u \biggl) du \biggl]  .
\end{equation}
Note that since $t_y(y) = t_y(y+2\pi)$ is a periodic function of $y$ and since 
$\int^{2 \pi}_0 t_y(y) dy = 0$, then the last exponential function in Eq.(6) 
has to be a periodic function of $x$ with a period $2\pi v_F/\omega_c$.
Therefore, the wave functions (6) can be rewritten in a form of the Fourier series,
\begin{equation}
\Psi^{\pm}_{\epsilon}(x,p_y,\sigma) =  \exp(\pm i p_F x) 
\exp \biggl( \pm i \frac{\delta \epsilon}{v_F}x \biggl)
\exp \biggl( \pm i \frac{\mu_B \sigma H}{v_F}x \biggl)  \nonumber\\ 
\sum^{+ \infty}_{n=- \infty} A_n (p_y) \exp \biggl(i \frac{\omega_c n}{v_F} x \biggl) .
\end{equation}
As it directly follows from Eq.(7), 2D electron spectrum (3) in a magnetic field (4), becomes pure 1D and corresponds to an infinite number of 1D FS, located near
$p_x \simeq p_F$ and $p_x \simeq - p_F$,
\begin{equation}
\delta \epsilon^{\pm} (p_x) = \pm v_F (p_x \mp p_F) + n \omega_c 
-  \mu_B \sigma H  ,
\end{equation}
where $n$ is an integer quantum number.
Electron spectrum (8) in shown Fig.1.

Note the a metallic phase with 1D spectrum (8) is unstable with respect to
the CDW phases formation because of its 1D "nesting" 
properties.  
Since the FICDW instability corresponds to a pairing of an electron near $p_F$ 
and a hole near $-p_F$ (and vice versa) with the same spins, then we expect that  possible projections along $x$-axis of the FICDW wave vectors are quantized 
at low enough temperatures (see Fig.1),
\begin{equation}
Q^n_x = 2p_F \pm 2 \mu_B H /v_F + n (\omega_c/v_F) , \ \ \pi T \leq \omega_c ,
\end{equation}
where the quantization of the electron spectrum (8) is important.
Therefore, at low temperatures, we expect a competition between 
the quantized FICDW order parameters (9) and have to choose the order 
parameter, corresponding to the highest 
transition 
temperature. 

Below, we consider a problem about a formation of the FICDW phases due
to electron-phonon interactions by means of the Feynman diagram 
technique [20,21].
In particular, we consider the FICDW order parameter in the following form,
\begin{equation}
\Delta(x,y) = \exp (iQ_x x) \exp (iQ_y y) 
+ c.c. \ ,  \ \ Q_x= 2p_F + q_x, \ Q_y = (\pi - 2 \alpha)/a_y + q_y 
\end{equation}
(where c.c. stands for a complex conjugated quantity), which allows to take into 
account deviations of the FICDW "nesting" vector from its ideal
value (2) both along $x$- and $y$-axis.
In a mean field approximation, a phase transition temperature between the metallic and FICDW phases is defined by the so-called electron polarization operator [20,21],
\begin{equation}
\frac{1}{g^2} = - \int^{2 \pi}_0 \frac{d (p_y a_y)}{2 \pi} \sum_{\sigma} T \sum_{\omega_n} \int^{+ \infty}_{-\infty}
dx_1 g^{\sigma}_{--} (i \omega_n;x,x_1;p_y-Q_y)
g^{\sigma}_{++}(i \omega_n;x_1,x;p_y) \exp[iq_x(x-x_1)] ,
\end{equation}
where $g$ is an electron-phonon coupling constant, $\omega_n$ is the 
Matsubara frequency.

Note that  Green functions of electrons near $p_F$ and $-p_F$, $g^{\sigma}_{++}(...)$ 
and $g^{\sigma}_{--}(...)$, respectively, can be determined from the corresponding electron wave 
functions (6) and spectrum (8) [21].
After substitution of the Green functions into Eq.(11) and some calculations, 
we obtain the following equations, which determine transition temperature
to the FICDW phases (10),
\begin{eqnarray}
&&T_{FICDW} \simeq \omega_c \exp \biggl[ -\frac{1}{g_{eff} (t'_y) g_{eff} (H)} \biggl] , 
\nonumber\\
&&g_{eff} (t'_y) = \frac{1}{2\ln(t'_y/t^*_y)} , \ \
g_{eff}(H) = MAX_{n,\delta t_y} \biggl<
\cos [\phi (x,p_y) +nx] \biggl>_{x,p_y}, \ \
\nonumber\\
&&\phi (x,p_y) = - \frac{4 \delta t_y}{\omega_c} \sin (x/2) \cos (p_y a_y) +
\frac{4 t'_y}{\omega_c} \sin(x) \cos(2p_y a_y) ,
\end{eqnarray}
with the quantized $x$-component of the wave vector,
\begin{equation}
q_x = \pm 2 \mu_B H /v_F + n (\omega_c/v_F) .
\end{equation}
[Here, $MAX_{n,\delta t_y}$ denotes a maximization procedure over
the integer quantum number $n$ and continuous variable $\delta t_y$, whereas
$<...>_{x,p_y}$ stands for an averaging procedure over the 
variables $x$ and $p_y$.]
Note that a metallic phase is supposed to be stable at $H=0$, which means 
that $t'_y > t^*_y$ in Eq.(12), where $t^*_y$ is a value of the parameter 
$t'_y$, corresponding to a CDW phase transition at $H=0$ 
and $T=0$.
The FICDW transition temperature (12) is calculated with the so-called logarithmic accuracy, where we use the following inequalities: 
$T \ll \omega_c$ and $t'_y \ll t_y$.

Eq.(12) and its numerical analysis are the main results of the Letter. 
The distinct feature of Eq.(12) is that the ratios of the FICDW magnetic critical fields 
(i.e., phase transition fields to the FICDW phases with different quantum numbers (13)) do not depend on any 
parameter. 
Numerical calculations of the effective coupling constant $g_{eff}(H)$ in Eq.(12)
are presented in Fig.2, where each FICDW phase is characterized by some 
quantum number $n$ in Eq. (13)
(see the figure caption).
The calculated ratios,  $H_1/H_0 = 0.73, \ H_2/H_0 = 0.59, \
H_3/H_0 = 0.49, \ H_4/H_0 = 0.42$ (where $H_n$ is a critical field
of a phase transition between the FICDW phases with numbers 
$n$ and $n+1$) are compared with the experimental data [18] 
in Table 1.
As it follows from the Table, there is an excellent agreement between the
calculated values $H_1/H_0$ and $H_2/H_0$ and the measured 
ones.
As to the measured ratio $H_3/H_0 \simeq 0.4$, it is in a satisfactorily agreement
with the corresponding calculated value, $H_3/H_0 = 0.49$.
On the other hand, we cannot exclude [22] that, in the experiments [18], in fact,
the fourth critical field, $H_4$, was measured instead of the third one, $H_3$.
This would give an excellent agreement with the corresponding calculated value, 
$H_4/H_0 = 0.42$.
Another important property of Eq.(12) is that the phase transition temperature 
is the same for two wave vectors, corresponding to signs (+) and (-) 
in Eq.(13).

In our opinion, a very good correspondence between the results of the present 
theory and the experimental data [14-18] is a strong argument in a favor of our 
model, based on electron-phonon 
interactions.
On the other hand, we point out that the previous simplified model [19] is not 
in a quantitative agreement with the existing 
experiments.
Indeed, we have numerically analyzed Eq.(11) of Ref.[19] and found that,
in the framework of the simplified model, $H_1/H_0 = 0.55$, $H_2/H_0=0.38$,
and $H_3/H_0=0.29$, which is in obvious disagreement with the experimental
data [18] (see Table 1).
Therefore, it is crucial to maximize the FICDW phase transition temperature 
(12) over two components of the wave vector (10), $q_x$ and $q_y$, 
which is not done in 
Ref. [19].
We note that the following inequalities, $T \ll \omega_c$ and $t'_y \ll t_y$, are
used for the derivation of Eqs.(12),(13).
Therefore, we do not take into account the finite temperature effects, described in 
Refs. [23,24] for the case of the FISDW phases.
The next step in our studies will be to suggest a relative universal theory of 
the FISDW phase diagram and to compare its results with the existing experimental 
data.
This problem will be considered in details elsewhere [25].

\begin{table}
\caption{\label{tab:table4} Theoretical and experimental [18] values of the critical fields 
ratios for different pressures.
}
\begin{ruledtabular}
\begin{tabular}{ccddd}
Critical fields&$H_1/H_0$&H_2/H_0&H_3/H_0&H_4/H_0\\
\hline
Theory
  &0.73&0.59&0.49&0.42\\
  P= 4 kbar
  &0.77&0.59&0.40&-\\
  P= 3.5 kbar&0.74&0.57&0.37&-\\
  P= 3 kbar&0.75&0.56&0.40&-\\
\end{tabular}
\end{ruledtabular}
\end{table}

One of us (A.G.L.) is thankful to N.N. Bagmet (Lebed), J.S Brooks, and M.V. Kartsovnik for very useful discussions.
This work was supported by the NSF grant DMR-0705986.

$^*$Also Landau Institute for Theoretical Physics,
2 Kosygina Street, Moscow, Russia.

\pagebreak

\begin{figure}[h]
\includegraphics[width=6.5in,clip]{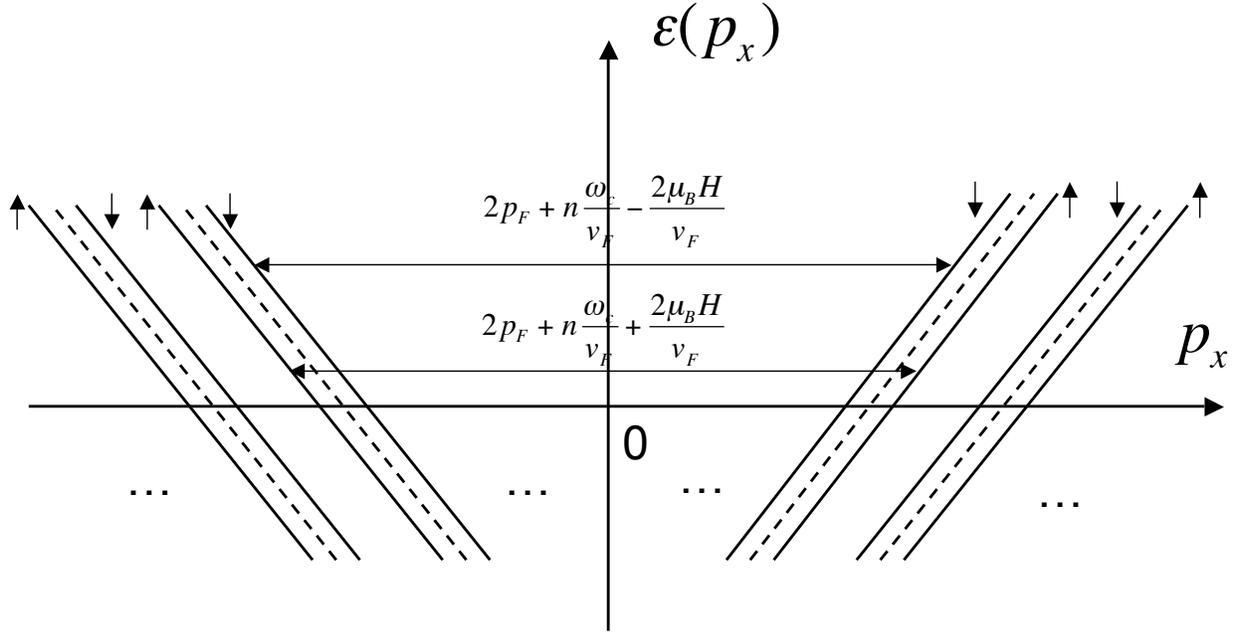}
\caption{
A schematic view of the quantized electron spectrum (8) near $p_x \simeq p_F$
and $p_x \simeq - p_F$.
There exist an infinite number of 1D Fermi surfaces, characterized by quantum
number $n$, with each of them being split due to an electron spin.
As a result, at low enough temperatures, there exist a competition between
infinite number of "nesting" vectors, corresponding to Eq.(9).
 }
\label{fig1}
\end{figure}

\begin{figure}[h]
\includegraphics[width=6.5in,clip]{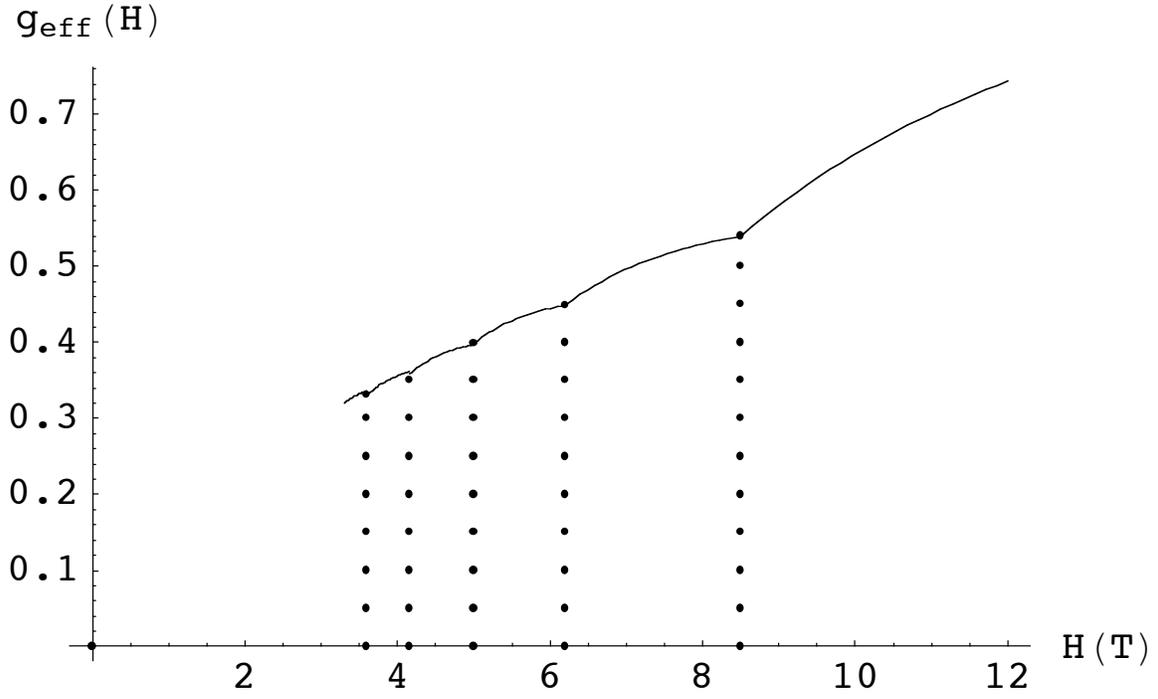}
\caption{
Numerically calculated effective coupling constant, $g_{eff}(H)$, which defines
the metal-FICDW phases transition temperature (see Eq.(12)), is shown by
a solid line.
Phase transitions between different FICDW phases, characterized by different
quantum numbers $n$ in Eq.(13), are shown by dotted lines.
Phase $n=0$ corresponds to $H > 8.5 \ T$, phase $n=1$ - $8.5 \ T > H > 6.2 \ T$,
phase $n=2$ - $6.2 \ T > H > 5 \ T$, phase $n=3$ - $5 \ T > H > 4.15 \ T$,
phase $n=4$ - $4.15 \ T > H > 3.6 \ T$, phase $n=5$ - $3.6 \ T > H$.
 }
\label{fig2}
\end{figure}

\pagebreak

\end{document}